\documentclass[sigconf,screen]{acmart}

\usepackage{algorithm}
\usepackage{graphicx}
\usepackage{textcomp}
\usepackage{xcolor}
\usepackage{cases, algcompatible, bm, multirow}
\usepackage{pifont}
\usepackage{booktabs}
\usepackage{xspace}
\usepackage{subfigure}
\usepackage{multirow}
\usepackage{threeparttable}

\newcommand{\we}{\textit{MARCA}\xspace}

\AtBeginDocument{%
  }

\copyrightyear{2024}
\acmYear{2024}
\setcopyright{acmlicensed}\acmConference[ICCAD '24]{IEEE/ACM International Conference on Computer-Aided Design}{October 27--31, 2024}{New York, NY, USA}
\acmBooktitle{IEEE/ACM International Conference on Computer-Aided Design (ICCAD '24), October 27--31, 2024, New York, NY, USA}
\acmDOI{10.1145/3676536.3676798}
\acmISBN{979-8-4007-1077-3/24/10}

\begin{document}

\title{MARCA: \underline{M}amba \underline{A}ccelerator with \underline{R}e\underline{C}onfigurable \underline{A}rchitecture}

\author{Jinhao Li$^{1*}$, Shan Huang$^{1*}$, Jiaming Xu$^{12}$, Jun Liu$^1$, Li Ding$^1$, Ningyi Xu$^1$, Guohao Dai$^{1\dag}$}

\affiliation{%
  \country{$^1$Shanghai Jiao Tong University, $^2$Infinigence-AI, $^*$Equal contributions} 
}

\affiliation{%
  \country{$^\dag$Corresponding author: daiguohao@sjtu.edu.cn} 
}

\renewcommand{\shortauthors}{Jinhao Li et al.}



\begin{abstract}
State space model (SSM) especially Mamba has demonstrated remarkable capabilities in various domains.
Compared to Transformers, Mamba reduces the quadratic computational complexity and achieves a higher algorithm accuracy (\textit{e.g.}, the accuracy of Mamba-2.8b is higher than OPT-6.7b). 
However, challenges still exist in accelerating Mamba computations.
\textbf{(1) Incompatibility between element-wise operations and Tensor Core.} 
Linear operations (matrix multiplications) and element-wise operations are the two dominating operations in Mamba.
The time proportion of element-wise operations escalates significantly (\textit{e.g.}, $>$60\% with 2048 input length). 
These operations do not need reduction, which is not compatible with the existing Tensor Core-based architectures (\textit{e.g.}, 1/16 normalized speed). 
\textbf{(2) Large area overhead for nonlinear function unit.} The optimized nonlinear function unit like exponential unit still occupies $>$30\% of the processing element (PE) area.
\textbf{(3) Large memory access but limited data sharing for element-wise operations.} 
Linear and element-wise operations in Mamba exhibit large compute intensity variance (\textit{e.g.}, $\sim$3 orders of magnitude) and large read/write ratio variance (\textit{e.g.}, $>$3 orders). 
Due to the limited data sharing in element-wise operations, it is useless to apply the existed methods like tiling to element-wise operations.

In response to these challenges, we propose a Mamba accelerator with reconfigurable architecture, \we.
Then, we propose three novel approaches in this paper.
\textbf{(1) Reduction alternative PE array architecture} for both linear and element-wise operations. 
For linear operations, the reduction tree connected to PE arrays is enabled and executes the reduction operation.
For element-wise operations, the reduction tree is disabled and the output bypasses.
\textbf{(2) Reusable nonlinear function unit} based on the reconfigurable PE. 
We decompose the exponential function into element-wise operations and a shift operation by a fast biased exponential algorithm,
and the activation function (SiLU) into a range detection and element-wise operations by a piecewise approximation algorithm.
Thus, the reconfigurable PEs are reused to execute nonlinear functions with negligible accuracy loss.
\textbf{(3) Intra-operation and inter-operation buffer management strategy}.
We propose intra-operation buffer management strategy to maximize input data sharing for linear operations within operations, and inter-operation strategy for element-wise operations between operations.
We conduct extensive experiments on Mamba model families with different sizes.
\we achieves up to 463.22$\times$/11.66$\times$ speedup and up to 9761.42$\times$/242.52$\times$ energy efficiency compared to Intel Xeon 8358P CPU and NVIDIA Tesla A100 GPU implementations, respectively.

\end{abstract}

\maketitle

\section{Introduction}
State space model (SSM) especially Mamba~\cite{mamba} has demonstrated remarkable capabilities in various domains including language, images~\cite{dosovitskiy2020image,zhu2024vision,peebles2023scalable}, audio~\cite{malik2021automatic}, and genomics~\cite{wang2024graph,min2022transformer}. 
Before Mamba, Transformer-based large language models~\cite{opt,llama} have achieved great success in sequence modeling due to the self-attention mechanism~\cite{attention}.
However, it suffers from handling the quadratic growth of storage and computational complexity as the sequence length increases and struggles to handle sequence with long range dependency.
Compared with the pioneers of SSMs~\cite{lssl,s4,dss_s4d} and Transformer, Mamba achieves higher accuracy in algorithm (\textit{e.g.}, the accuracy of Mamba-2.8b is higher than OPT-6.7b~\cite{opt}) and more efficient computation in hardware.
Therefore, Mamba, as a foundation model~\cite{foundation_model}, has been applied in various domains including vision~\cite{zhu2024vision,patro2024simba}, graphics~\cite{li2024stg,wang2024graph,behrouz2024graph}, medical~\cite{yue2024medmamba}, and point cloud~\cite{liu2024point,zhang2024point,liang2024pointmamba}.

\begin{figure}[!t]
  \centering
  \vspace{-5pt}
  \includegraphics[width=0.4\textwidth]{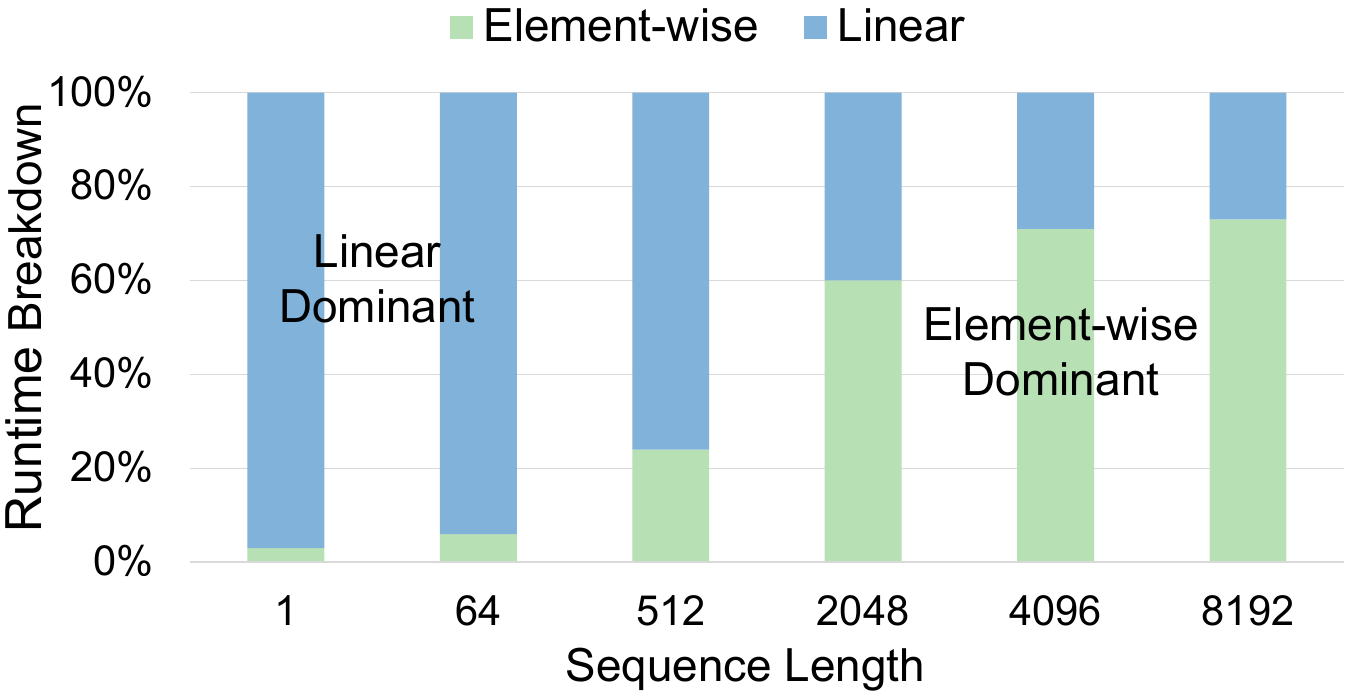}
  \vspace{-12pt}
  \caption{Runtime breakdown with different sequence lengths in Mamba. Element-wise operations contribute a large fraction of the runtime with long sequence length while linear operations are dominant with short length.}
  \vspace{-18pt}
  \label{fig:time_proportion}
\end{figure}

 \begin{figure*}[!t]
  \centering
  \vspace{-5pt}
  \includegraphics[width=0.98\textwidth]{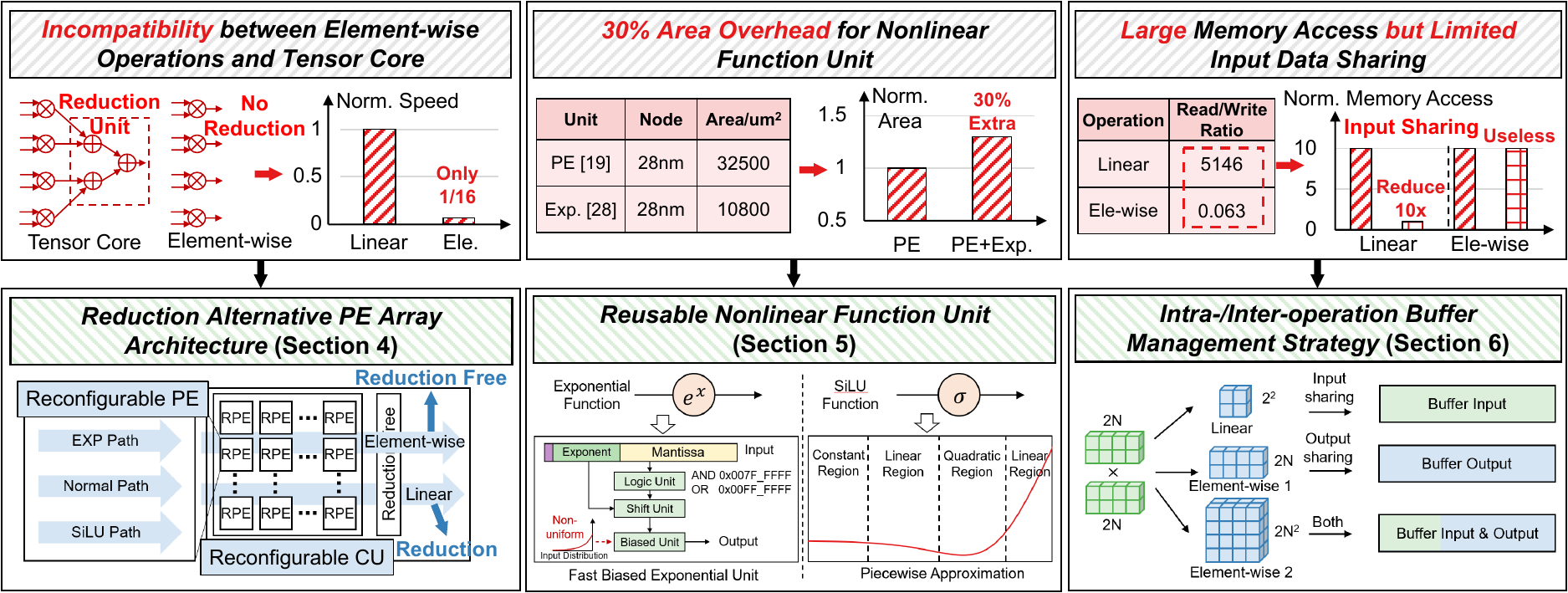}
  \vspace{-12pt}
  \caption{Challenges in Mamba computation: (1) incompatibility between element-wise operations and Tensor Core, (2) 30\% area overhead for nonlinear function unit, and (3) large memory access but limited input data sharing for element-wise operations. We propose three novel contributions in \we: (1) reduction alternative PE array architecture, (2) reusable nonlinear function unit, and (3) intra-operation and inter-operation buffer management strategy, to solve these challenges.}
  \vspace{-15pt}
  \label{fig:overview}
\end{figure*}

However, there is limited research on optimizing for Mamba processing.
Therefore, we profile the processing carefully and identify three main challenges in Mamba computations:
\textbf{(1) Incompatibility between element-wise operations and Tensor Core.}
Linear operations (\textit{e.g.}, matrix multiplication and convolution) and element-wise operations are two dominating operations in Mamba.
Tensor Core~\cite{tensorcore_a100,tensorcore_h100} is a typical domain specific architecture with reduction tree focused on accelerating these linear operations with high the compute intensity (\textit{e.g.}, $>$1000 FLOPs/Byte).
However, as shown in Figure~\ref{fig:time_proportion}, the time proportion of element-wise operations escalates significantly as sequence length increases (\textit{e.g.}, $>$60\% with 2048 input length).
Because the element-wise operations do not need reduction, applying them on Tensor Core-based architecture should introduce large amount of invalid computations, leading to an extreme inefficiency (\textit{e.g.}, 1/16 normalized speed). 
\textbf{(2) Large area overhead for nonlinear function unit.}
Exponential function and Sigmoid Linear Unit (SiLU) function are two nonlinear functions in Mamba. Previous methods~\cite{exp_unit,nilsson2014hardware,reggiani2023flex} often design specific unit to compute them, leading to much more area overheads. 
As shown in Figure~\ref{fig:overview} middle bottom, the optimized nonlinear function unit such exponential function still occupy 30\% of the PE area~\cite{tstc,exp_unit}.
\textbf{(3) Large memory access but limited input data sharing for element-wise operations.}
Linear and element-wise operations in Mamba exhibit large compute intensity variance (\textit{e.g.}, $\sim$3 orders of magnitude) and large read/write ratio variance (\textit{e.g.}, $>$3 orders of magnitude). 
Due to the limited data sharing in element-wise operations, it is useless to apply the existed methods like tiling~\cite{li2019coordinated} to element-wise operations.

In response to these challenges, we propose \we, a Mamba accelerator with reconfigurable architecture, to support fast and energy-efficient Mamba computations. 
Our contributions are as follows:
    
\textbf{(1) Reduction alternative PE array architecture} for both linear and element-wise operations.
For linear operations, the reduction tree connected to PE arrays is enabled and executes the reduction operation.
For element-wise operations, the reduction tree is disabled and bypasses the results.
    

\textbf{(2) Reusable nonlinear function unit} based on reconfigurable PE arrays.
We decompose the exponential function into element-wise operations and a shift operation by a fast biased exponential algorithm.
 We also decompose the activation function (SiLU) into a range detection and element-wise operations by a piecewise approximation algorithm.
Thus, the reconfigurable PEs are fully reused to execute nonlinear functions with negligible accuracy loss.
    
\textbf{(3) Intra-operation and inter operation buffer management strategy.}
We propose intra-operation and inter-operation buffer management strategy for linear and element-wise operations. 
For linear operations, the buffer pool is managed as an input buffer to maximize the input data sharing within each operation. 
For element-wise operations, the buffer pool is managed as an output buffer to maximize the output data sharing between operations. 


We implement \we in \textit{Verilog} and design a cycle-accurate simulator to evaluate \we. 
We conduct extensive experiments on Mamba with different model sizes (\textit{i.e.}, Mamba-130M to Mamba-2.8B).
As a result, \we achieves up to 463.22$\times$/11.66$\times$ speedup and up to 9761.42$\times$/242.52$\times$ energy efficiency compared to Intel Xeon 8358P CPU and NVIDIA Tesla A100 GPU implementations, respectively. 



\section{Background}\label{sec:background}

\subsection{State Space Model}\label{sec:background:ssm}

The continuous state space model in equation~\ref{eq:cont_ssm} defines a linear mapping from an input signal $x(t)\in \mathbb{R}^{M}$ (a function of time $t$) to output signal $y(t)\in \mathbb{R}^{M}$ through a hidden state $h(t) \in \mathbb{R}^{N}$:
\begin{equation}\label{eq:cont_ssm}
\begin{split}
  h'(t) &= A(t)h(t)+B(t)x(t) \\
   y(t) &= C(t)h(t)
\end{split}
\end{equation}
where state matrix $A(t) \in \mathbb{R}^{N\times N}$, input matrix $B(t) \in \mathbb{R}^{N\times M}$, output matrix $C(t) \in \mathbb{R}^{M\times N}$, and the change of hidden state $h'(t)\in \mathbb{R}^{N}$.
Classical SSM has dynamic parameters (\textit{e.g.}, $A,B,C$) that change over time. 
However, when they are constant the dynamics are invariant through time, which is known as a linear time-invariant (LTI) system~\cite{lti}, and is equivalent to a (continuous) convolution.

Data like words and tokens in the real world is discrete instead of continuous, so equation~\ref{eq:cont_ssm} must be discretized to be applied to an sampled input sequence $x=(x_0,x_1,x_2,...)$ instead of continuous function $x(t)$. 
An additional step size parameter $\Delta$ is required that represents the resolution of the input. 
Conceptually, the inputs $x(n)$ can be viewed as uniformly-spaced samples from an implicit underlying continuous signal $x(t)$, where $x(k)=x(k\Delta)$.
The discretization process from continuous time signal processing to discrete processing is the fact that the SSM has equivalent forms of solving partial differential equations.
To illustrate discretization, the simplest method is to apply Taylor series method~\cite{taylor} which turns the equation~\ref{eq:cont_ssm} into the first-order approximation:
\begin{equation}\label{eq:ssm}
\begin{split}
  h_n &= h_{n-1}+\Delta(Ah_{n-1}+Bx_{n}) \\
      &= (I+\Delta A)h_{n-1}+(\Delta B)x_{n} \\
      &= \overline{A}h_{n-1}+\overline{B}x_n \\
  y_n &= Ch_n
\end{split}
\end{equation}
where the discrete version of parameters has the same shapes as the original continuous version: $A \in \mathbb{R}^{N\times N}$ and $B \in \mathbb{R}^{N\times M}$. The discretized system depends on $\Delta$ to generate $\Delta A$ and $\Delta B$. 
Therefore, $\Delta$ can be interpreted as parameters that modulates the SSM instead of as a fixed step size. 

\begin{figure}[!t]
  \centering
  \includegraphics[width=0.47\textwidth]{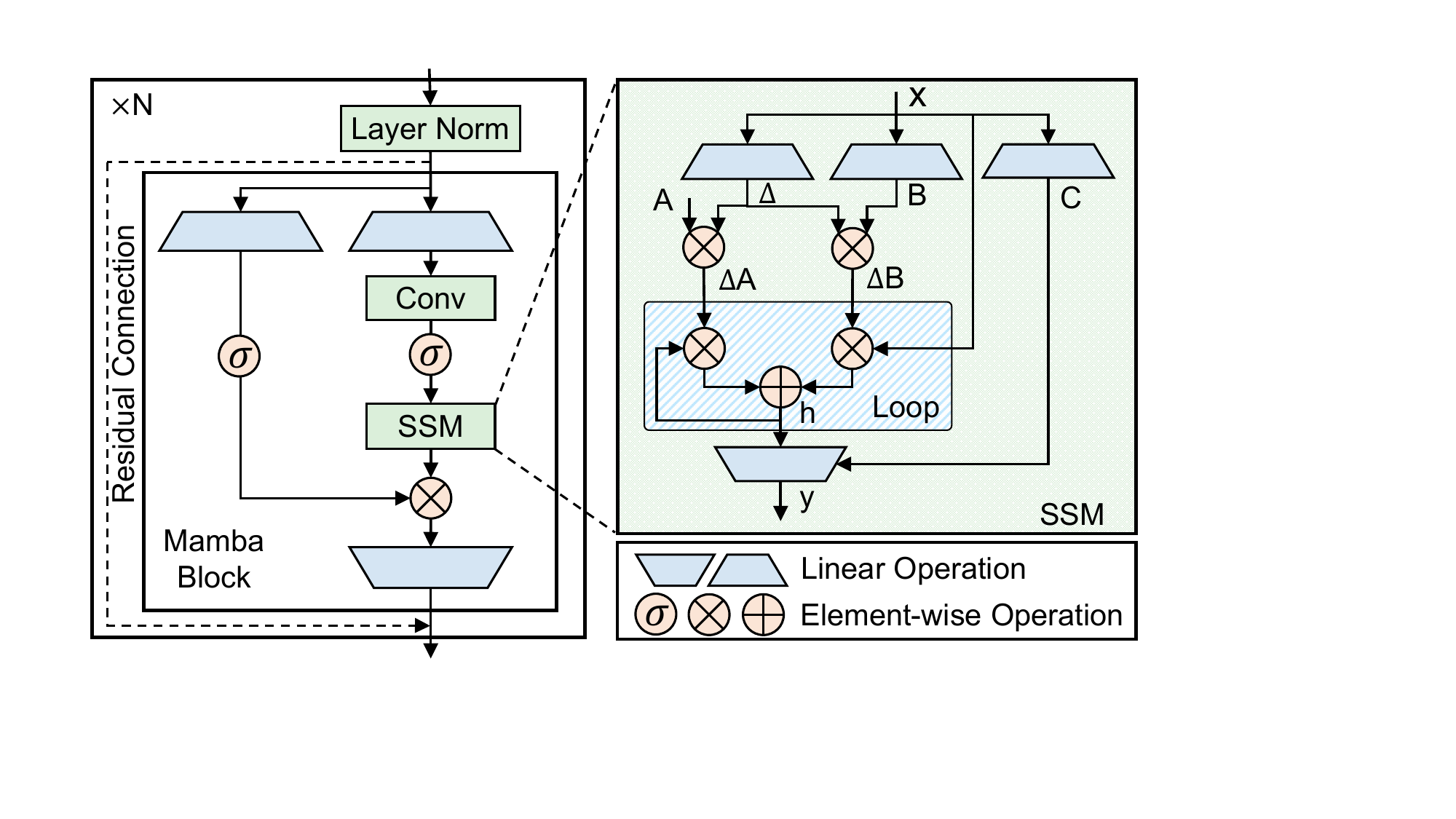}
  \vspace{-16pt}
  \caption{Computational flow in Mamba block and SSM. Mamba model consists of N blocks with residual connection. In SSM, $\Delta$, $B$ and $C$ are generated by input $x$. Then it performs loops to update hidden state $h$, and generates output $y$.}
  \vspace{-18pt}
  \label{fig:mamba_block}
\end{figure}

\subsection{Mamba}\label{sec:background:mamba}

Mamba~\cite{mamba} introduces selective mechanism into SSM and proposes an implementation of selective state space model layer.
Mamba block consists of a layer normalization, several linear projections, a convolution, a SSM block, and a residual connection.
In each layer, the input sequence is first processed by a linear projection and then processed by the convolution.
Then it is processed by an activation and then processed by the SSM. 
After SSM, the main branch is multiplied by the collateral branch including a linear projection and an activation (\textit{e.g.}, SiLU~\cite{silu}) to generate the combined result.
After combination, the result is processed by a linear projection.
Last, the result is added back to the input through the residual connection~\cite{resnet}. 
Instead of interleaving Mamba block and Feed-forward block, Mamba simply repeats the Mamba block homogenously. 

During SSM processing, the input $x$ undergoes a series of transformations. 
Firstly, it is subjected to three linear projections, resulting in $\Delta$, $B$ and $C$. 
Subsequently, $\Delta$ is involved in the Einstein summation operations~\cite{einsum} with $A$ and $B$ separately, generating $\Delta A$ and $\Delta B$. 
These intermediate results $\Delta A$ and $\Delta B$ are then multiplied with the hidden state and input $x$ by element-wise Einstein summation, respectively. 
After $L$ (sequence length) times iterations, the hidden state is updated for $L$ times. 
The outcomes of these operations are then combined through addition, resulting in the updated hidden state. 
Finally, the updated hidden state undergoes matrix multiplication with $C$, followed by a linear transformation, to generate the output. The whole computational flow in Mamba block is illustrated in Figure \ref{fig:mamba_block}.


\section{Architecture Overview}

\begin{figure*}[!t]
  \centering
  \vspace{-5pt}
  \includegraphics[width=0.98\textwidth]{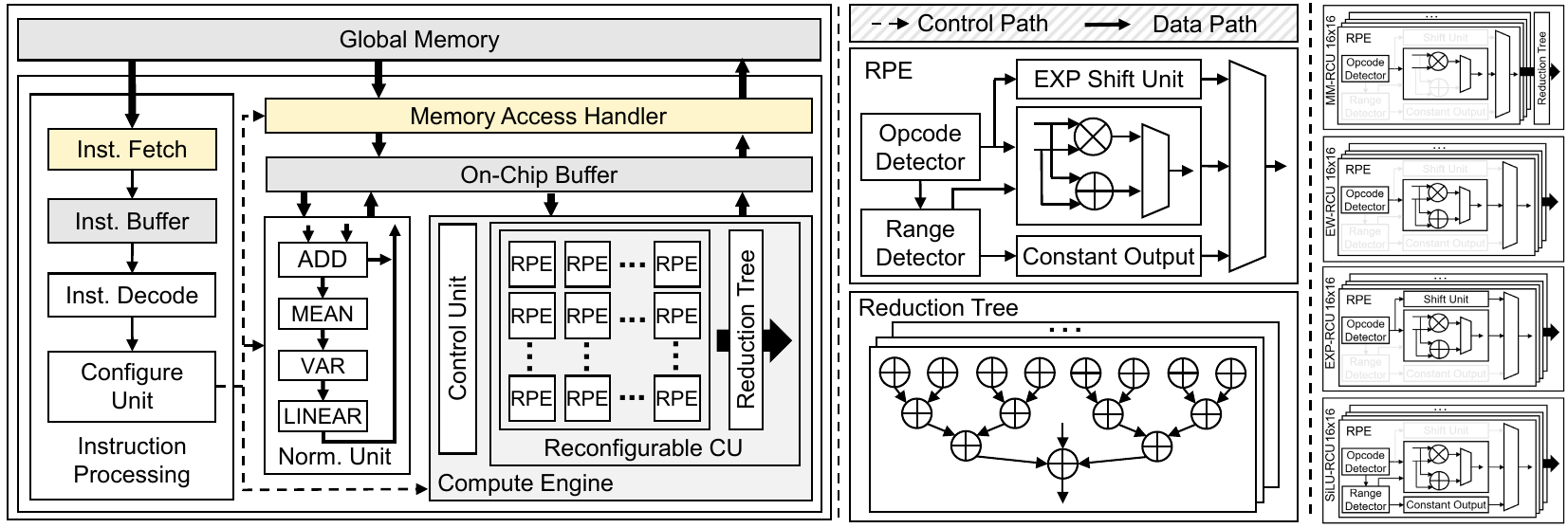}
  \vspace{-12pt}
  \caption{Left: Architecture of \we accelerator. \we mainly consists of an instruction processing, a normalization unit, an on-chip buffer,and a compute engine. Middle: Architecture of reconfigurable processing element and reduction tree in RCU. Right: Four reconfigurable modes of RCU, MM-RCU, EW-RCU, EXP-RCU, and SiLU-RCU.}
  \vspace{-10pt}
  \label{fig:architecture}
\end{figure*}

We propose a Mamba accelerator with reconfigurable architecture, \we, with reconfigurable computing units (CUs) and processing elements (PEs).
\we is a reconfigurable architecture whose instructions are all 64-bit, and contains 16 32-bit general-purpose Registers (Regs) and 16 32-bit Constant Registers (CRegs).
\we consists of four main parts: instruction processing, normalization unit, on-chip buffer, and computing engine, as depicted in Figure.~\ref{fig:architecture} left.

\begin{figure}[!t]
  \centering
  \includegraphics[width=0.48\textwidth]{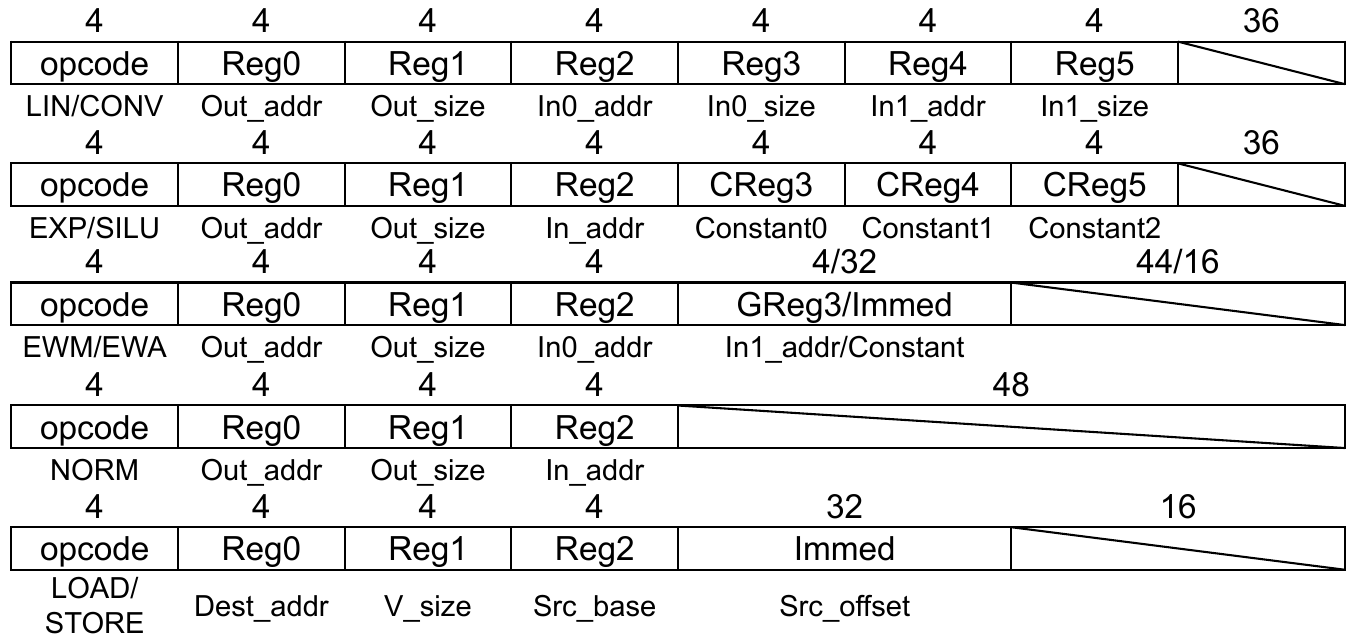}
  \vspace{-18pt}
  \caption{Instruction set architecture with 16 32-bit general-purpose registers and 16 32-bit constant registers. All instructions are 64-bit.}
  \vspace{-15pt}
  \label{fig:isa}
\end{figure}

\textbf{Instruction Processing.}
The instruction processing consists of two parts: instruction fetch and instruction decode. 
The instruction fetch unit fetches instructions from global memory and stores them in the instruction buffer. 
Then, the instruction decode unit reads instructions sequentially from the buffer and decodes them.
As shown in Figure~\ref{fig:isa}, the instruction set architecture (ISA) includes linear (\texttt{LIN}), convolution (\texttt{CONV}), normalization (\texttt{NORM}), element-wise multiplication (\texttt{EWM}), element-wise addition (\texttt{EWA}), exponential function (\texttt{EXP}), and SiLU (\texttt{SILU}). 
And \we also provides \texttt{LOAD} and \texttt{STORE} instructions to support moving data between global memory and on-chip buffer.
After decoding, the instructions are passed through the configure unit to pass configuration information to the following modules.


\textbf{Normalization Unit.}
The layer normalization is a important component that stabilizes range of intermediate values by normalizing layer activation. 
The normalization unit is responsible for computing the mean and variance of the data. 
The data is first summarized and accumulated to get the mean and then the variance is calculated.
Then, data undergoes a linear unit to obtain the normalized result.

\textbf{Compute Engine.}
The compute engine is responsible for linear, convolutional, and element-wise computations. 
It comprises a control unit and a reconfigurable compute units (RCUs). 
The control unit receives configuration information from configure unit and fetches data from on-chip buffer to RCU.
When the computation is completed, it notifies the instruction processing pipeline to continue decoding and executing the next instruction.


\section{Reduction Alternative PE Array Architecture}\label{sec:method1}

\subsection{Challenge}

Linear operation and element-wise operation are two dominating operations in Mamba.
Linear operations like matrix multiplications are usually accelerated by Tensor Core, which is a domain specific architecture with the reduction tree.
Due to the reduction of partial inner product, the linear operations exhibit extremely high compute intensity (\textit{e.g.}, $>$1000 FLOPs/Byte).
However, as shown in Figure~\ref{fig:time_proportion}, the time proportion of element-wise operations escalates significantly as sequence length increases (\textit{e.g.}, $>$60\% with 2048 input length).
Because the element-wise operations do not need reduction, applying them on Tensor Core-based architecture should introduce large amount of invalid computations, leading to an extreme inefficiency (\textit{e.g.}, 1/16 normalized speed). 
We call it the incompatibility between element-wise operations and Tensor Core.

\subsection{Motivation and Insights}

Our motivation stems from the inefficiencies observed in element-wise computations on tensor cores. 
Because the attention operations contain non-linear softmax, previous Transformer accelerators~\cite{wang2022via,lu2020hardware} employed independent hardware units to support both attention and linear operations. 
In contrast, Mamba architecture only consists of linear operations, element-wise operations, and a few activation functions. 
The only difference of linear of element-wise operations is whether execute reduction or not. 
Recognizing this simplicity, \textbf{our key insight is that by disabling the reduction tree, the Tensor Core-based PE arrays can execute element-wise operations.}

\subsection{Approach}
\textbf{Reduction Alternative PE Arrays.}
The reduction alternative processing element (PE) arrays support configurablility. 
The reduction tree consists of 16 slices (taking 16 for instance) and operates in two modes: reduction mode for linear operations and non-reduction mode for element-wise operations. 
In reduction mode, the reduction tree is enabled.
For a 16-to-1 reduction tree slice, the outputs from 16$\times$1 PE arrays are fed into one reduction tree slice, which employs multi-level additions to compute the sum.
In non-reduction mode, the reduction tree is enabled and the outputs from 16$\times$16 PE arrays skip the reduction directly. 
In addition, the last-level addition in each slice supports three input to accumulate the partial results for linear operations.

We define that a reconfigurable computing unit (RCU) consists of  16$\times$16 PE arrays and a reduction tree, as shown in Figure~\ref{fig:architecture} left. 
Each reconfigurable PE (RPE) is configured to support three main computations with three data paths: a shift path for exponential function, a piecewise path for SiLU function, and a normal path for addition or multiplication. 
The normal path contains a floating-point multiplier, a floating-point adder, and a multiplexer unit to handle element-wise multiplication and addition computations.

\textbf{Reconfigurable Computing Unit.}
We provide a detailed explanation of how the RCU operates for the four specific computations in Mamba, as shown in Figure~\ref{fig:architecture} right.

\textit{\underline{MM-RCU.}}
The RCU is configured as a matrix multiplication mode (MM-RCU) to support linear operations. 
The reduction tree in RCU and the floating-point multiplier units in RPE are enabled. 
Therefore, for a matrix multiplication operation of two matrices of size 16$\times$16, the results calculated by the 16$\times$16 array of multipliers are then passed through the reduction tree to produce 16 final results. 
Then, this process is repeated totally 16 times to obtain the complete result of the 16$\times$16 matrix multiplication. 
To support the accumulation of partial sums, an additional adder is added at the final level of the reduction tree.

\textit{\underline{EW-RCU.}}
When RCU is configured as element-wise mode (EW-RCU), the reduction tree is disabled, while the P2D buffer is enabled, and the floating-point multiplier or adder units in RPE are activated. 
For an element-wise multiplication operation of two 16x16 matrices, the results calculated by the 16$\times$16 array of RPEs maintains the same dimensions of 16$\times$16 and are output to the buffer in parallel.

\textit{\underline{EXP-RCU.}}
When RCU is configured as exponential mode (EXP-RCU), the reduction tree is disabled. 
The floating-point multiplier, adder units, and exponential shift unit in RPE are enabled. 
Therefore, for a 16$\times$16 matrix performing exponential function operation, the RCU first executes element-wise multiplication by using the multipliers, then executes element-wise addition by using the adders. 
Afterward, the exponential shift unit performs a logic operation, a shift operation and a biased operation to obtain the final output, as shown in Figure~\ref{fig:overview} right bottom.

\textit{\underline{SiLU-RCU.}}
When RCU is configured as SiLU mode (SiLU-RCU), the floating-point multiplier or adder, the range detector, and the constant output unit in RPE are enabled while the reduction tree is disabled.
For a 16$\times$16 input matrix, each input is first distinguished by range, and then depending on the difference of range, each input is processed by employing either the constant output or normal element-wise computations.
The SiLU-RCU decomposes the SiLU operation into 0, 2, or 4 instances of element-wise operations according to equation~\ref{eq:silu_approx}.

\section{Reusable Nonlinear Function Unit}\label{sec:method3}

\subsection{Challenge}
Exponential function and Sigmoid Linear Unit (SiLU) function are two nonlinear functions in Mamba. 
Previous methods~\cite{exp_unit,nilsson2014hardware,reggiani2023flex} often design specific unit based on lookup-table or Taylor series approximation to optimize these nonlinear computations.
However, the optimized nonlinear function unit such exponential function still occupy 30\% of the PE area~\cite{exp_unit,tstc}, leading to much more area overheads.

\subsection{Motivation and Insights}
A common approach is to utilize approximation functions to approximate exponential operations, thereby degrading exponential computations to quadratic or even linear operations.
Using linear approximation methods results in large precision loss ($>$4\% for Mamba-2.8b) while employing higher-order polynomial approximation leads to increased computational overhead.
We profile the range of these nonlinear functions.
The input for an exponential function is mostly between -7 and 0, especially for values slightly less than 0,
while the input range of the SiLU function is from -5 to 4.
On one hand, we can only approximate nonlinear functions only in these range to concentrate on preventing accuracy loss.
On the other hand, exponential function and SiLU are similar with element-wise operations except for scaling for each value. 
Therefore, to avoid increasing area overhead and prevent accuracy loss, our key insight is that \textbf{only by approximating these nonlinear functions in a small range, we can decompose them into several element-wise operations to reuse the reduction alternative PE array architecture in Section~\ref{sec:method1}.}

\subsection{Approach}

\textbf{Fast Biased Exponential Algorithm.}
Given the peculiar distribution observed in the inputs of the exponential function, namely the outer product of $\Delta$ and A, we leverage a set of data points $x=\frac{-7}{n}, n=1,2,...,200$ where the density increases as they approach zero to evaluate the deviation of the approximate calculation from the origin exponential value. Consequently, we modified the fast exp algorithm~\cite{fast_exp} to accommodate specific data ranges and appended a bias at the end to enhance precision and consists of three main steps:

\begin{enumerate}
    \item The input \( x \) is linearly transformed into \( x' \).
    \item \( x' \) is multiplied by $2^{23}$ then cast to an unsigned integer.
    \item View the x as float-point number and add the bias \( c \).
\end{enumerate}
 The fast exp algorithm aims to put $\frac{x}{ln2}$ into exponential bit of $e^{x}$ so that $e^{x} = 2^{\left(\frac{x}{ln2}\right)}$. Float data does not have a direct intuitive representation like unsigned integers, hence cannot be directly used as the exponent bits for $e^{x}$; therefore, conversion to uint is necessary. Additionally, a bias is required to compensate for conversion loss. The exponent of a float is typically subtracted by 127 during actual computations so we also need consider that. Now we can determine the coefficient \(a = \frac{1}{\ln(2)}\) and the term \(b = 127 + bias\), while \(c\) serves as the final bias, effectively reducing the average loss caused by specific data distributions.

\begin{figure}[!t]
    \centering
    \includegraphics[width=0.48\textwidth]{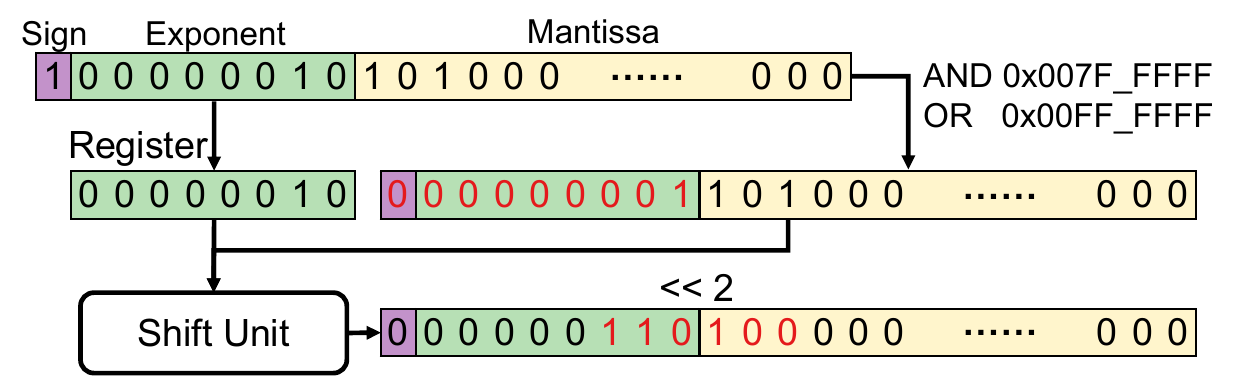}
    \vspace{-20pt}
    \caption{Hardware implementation of exponential shift unit for our fast exponential approximation.}
    \vspace{-18pt}
    \label{fig:approximation}
\end{figure}

\textbf{Piecewise SiLU Algorithm.}
The computational formula for the SiLU function is $SiLU(x) = x\cdot \sigma(x) = \frac{x}{1 + e^{-x}}$
which involves not only exponentiation but also division operations. Though employing the fast exp algorithm for approximation yields excellent precision, it introduces overhead due to the need for dividers, consequently impacting performance. Our preliminary profiling results indicate that the inputs to the SiLU function are mostly concentrated in the range [-5, 4]. Therefore, we propose segmenting the SiLU function within this interval for approximation. Increasing the number of segments enhances precision but also introduces more conditional branches, thus impacting performance. Consequently, we strike a balance between precision and performance and utilize a 4-segment piecewise function to approximate the SiLU function which is expressed by equation \ref{eq:silu_approx}: Within each interval, further piece-wise approximation is performed based on the distribution of the data.

\begin{equation}\label{eq:silu_approx}
f(x) = \left\{
\begin{aligned}
    &-0.0135, && \text{if } x < -5 \\
    &-0.06244x - 0.3457, && \text{if } -5 \leq x < -1.5 \\
    &0.232(x + 1.181)^{2} - 0.275, && \text{if } -1.5 \leq x \leq 0.75 \\
    &1.05x - 0.2781, && \text{if } x > 0.75
\end{aligned}
\right.
\end{equation}

\textbf{Reusable Nonlinear Function Unit.}
Based on two algorithms for nonlinear functions, we decompose the exponential function into element-wise operations and an shift operation, and SiLU into a range detection and element-wise operations.
Therefore, we only add a few logics and resue PE arrays to support nonlinear  functions.

\textit{\underline{As Exponential Function Unit.}} The original computation process of fast exp is overly complex, particularly the float-to-uint conversion unit incurring significant overhead, we have devised a dedicated conversion unit shown in Figure \ref{fig:approximation} for fast exp. Within this unit, the input \( x \) undergoes floating-point linear computation, resulting in \( x' \). Subsequently, we extract the 8 exponent bits of \( x' \), directly representing the required shift length—positive for left shifts and negative for right shifts. Next, employing logical operations, we convert the original floating-point number into the actual representation of the mantissa. Finally, the result is fed into a shift unit and adjusted through a bias unit to yield the output. Due to the ability to simplify the series of linear transformations applied to the input into a single linear transformation, the actual computation only requires 4 cycles.

\textit{\underline{As SiLU Unit.}} The SiLU unit introduces a range detector in the PE unit, responsible for determining the input's interval and executing the corresponding computation. Due to the most complex operation being quadratic, the actual computational overhead is minimal.


\begin{figure}[!t]
  \centering
  \vspace{-5pt}
  \includegraphics[width=0.48\textwidth]{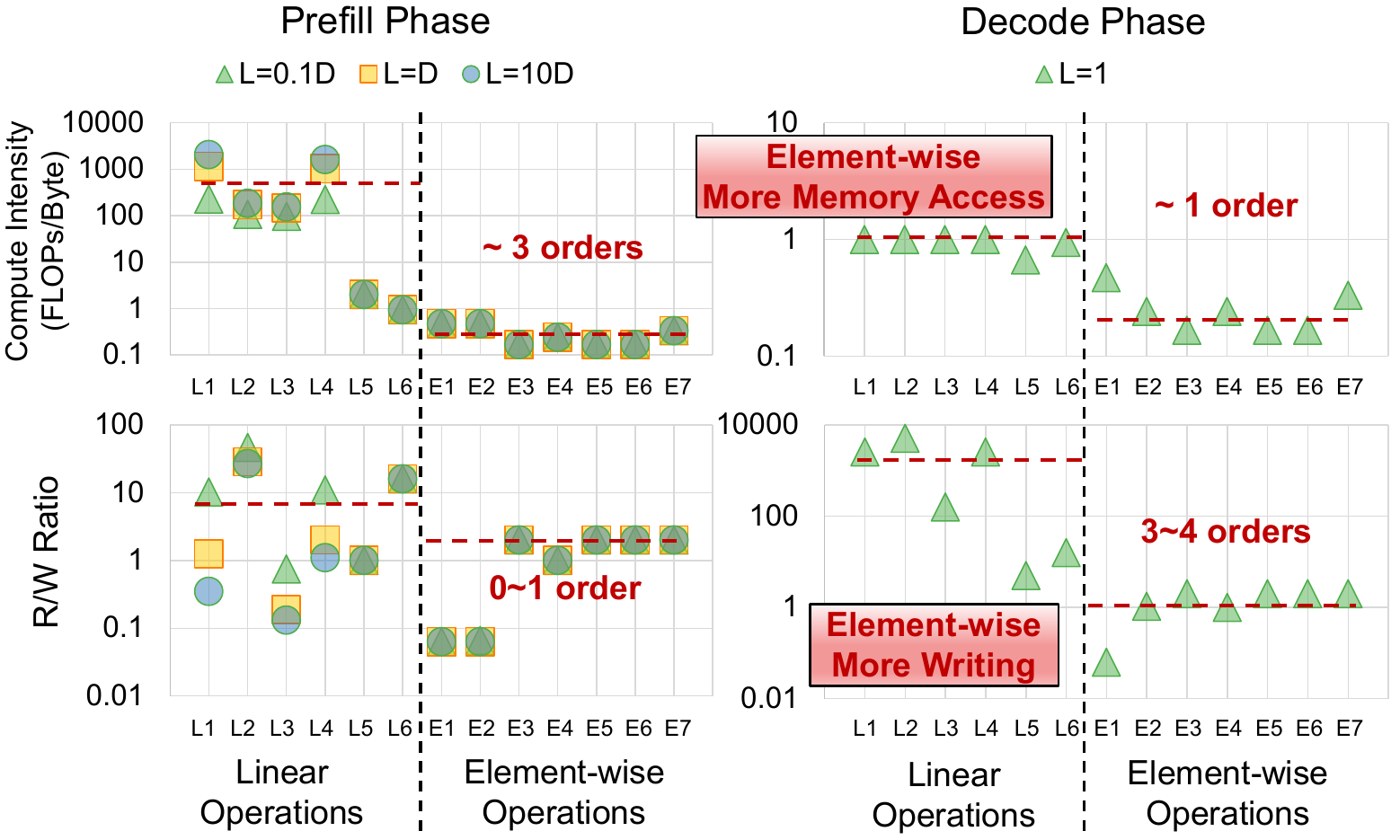}
  \vspace{-20pt}
  \caption{The difference of compute intensity and read/write ratio with different sequence length input in Mamba. Compared with linear operations, element-wise operations need more memory access and more memory writing.}
  \vspace{-20pt}
  \label{fig:read-write-ratio}
\end{figure}

\section{Intra-/Inter-Operation Buffer Management Strategy}\label{sec:method2}

\subsection{Challenge}
Linear and element-wise operations are two dominant operations in Mamba processing.
The feature of memory access for these two kinds shows three typical paradigms, that is, reading $2\times2N$ data and writing $2\times2$, $2N$, and $2N^{2}$ corresponding to linear projection (Linear), element-wise addition or multiplication (Element-wise 1), and element-wise outer product (Element-wise 2), as shown in Figure~\ref{fig:overview} right bottom. 
As shown in Figure~\ref{fig:read-write-ratio}, we depict a detailed breakdown of the memory read/write ratio with different input sequence length for various operations during Mamba computation process. 
It shows that the read/write ratio of the three operations differs by more than three orders of magnitude.
For linear operations with more input and less output, existing optimization methods primarily focus on tiling~\cite{li2019coordinated} and load the tiled input to on-chip buffer to maximize data sharing.
However, due to the computational characteristic of element-wise operations with more output and less input, it is redundant to apply input sharing methods like tiling.

\subsection{Motivation and Insights}
The computational characteristics of linear operations require reduction and input sharing, leading to a higher read/write ratio, which is suitable for existing input data sharing methods during each operation processing.
By reviewing Figure~\ref{fig:mamba_block}, we find that the element-wise operations are closely spaced within the SSM computation process.
And during the SSM process, the outputs of element-wise operations such as $\Delta A$, $\Delta B$, and $h$ are accessed repeatedly. 
Storing these output of element-wise operations on the on-chip buffer can significantly reduce the overhead of memory access for the next operation.
Therefore, our key insight is to \textbf{adjust different buffer management strategies for different operation types to maximize the data sharing with intra-operation and inter-operation.}

\begin{figure}[!t]
  \centering
  \vspace{-5pt}
  \includegraphics[width=0.48\textwidth]{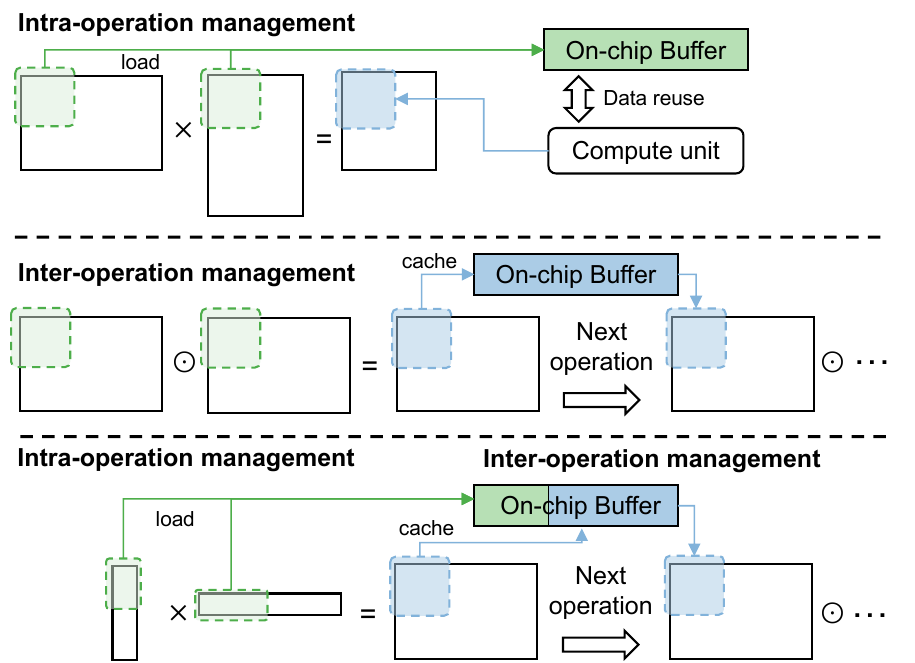}
  \vspace{-20pt}
  \caption{MARCA buffer management strategy for different operations.}
  \vspace{-20pt}
  \label{fig:buffer_management}
\end{figure}

\subsection{Approach}

Our operation-wise buffer management (BM) strategies contain intra-operation and inter-operation methods as shown in Figure ~\ref{fig:buffer_management}. 
Intra-operation buffer management (Intra-BM) is used to discover and manage data sharing within individual operations, while inter-operation buffer management (Inter-BM) is used to discover and manage data sharing between operations.

\textbf{Intra-operation Management.}
For linear operations, the whole on-chip buffers are configured as read buffers. The memory access handler loads linear inputs from global memory to fill the buffer. The computing unit then sequentially reads the required data from the buffer, performs calculations, and writes the computed results back to global memory.
For element-wise 2 operation, even though the read/write ratio is relatively low, it can be effectively regarded as a matrix multiplication with two reduction dimensions of size 1, hence requiring data sharing among inputs. Therefore, we reserve a small fraction region of the on-chip buffer to store them.

\textbf{Inter-operation Management.}
For element-wise 1, there is no data reuse in the computation of input data. 
The basic approach involves reading from global memory and directly writing back after computation. For individual computations, data reuse optimization like tiling is not feasible.
For element-wise 1 and 2 operations, to maximize buffer utilization, we primarily optimize data sharing among adjacent element-wise operations. 
As shown in equation~\ref{eq:ssm}, the update of hidden state $h$ is obtained by adding the product of $\Delta A$ and $h$ with the product of $\Delta B$ and $x$. Therefore, during the continuous updating process of state $h$, $h$ needs to be read and written $L$ times repeatedly. Additionally, the corresponding $\Delta A$, $\Delta B$, and $x$ need to be read $L$ times repeatedly. Hence, for element-wise operations concentrated in the SSM process, we cache the above immediate result in the buffer.

Through an operation-wise buffer management strategy, our method maximizes the reduction in memory access and minimizes idle computational resources, thereby accelerating the Mamba computation process.

\section{Experimental Results}\label{sec:experiment}
\subsection{Experimental Setup}
\textbf{Methodology.}
The performance and energy of \we are measured by using the following tools.

\textit{\underline{Architecture Simulator.}}
We design and implement a cycle-accurate simulator to measure execution time in number of cycles. 
This simulator models the microarchitectural behaviors of each module, which is integrated with Ramulator 2.0~\cite{ramulator} to simulate the behaviors of memory accesses to High Bandwidth Memory (HBM).

\textit{\underline{CAD Tools.}}
We implement and synthesize our design in Verilog to measure area, power, and critical path delay (in cycles) for each module. 
We use the Synopsys Design Compiler with the TSMC 28 nm standard VT library for the synthesis, and estimate the power using Synopsys PrimeTime PX.
The slowest module has a critical path delay of 0.9 ns including the setup and hold time, putting the \we comfortably at 1 GHz clock frequency.

\textit{\underline{Memory Measurements.}} 
The area, power, and access latency of the on-chip scratchpad memory are estimated using Cacti 7.0~\cite{cacti}. 
Since Cacti only supports down to 32 nm technologies, we apply four different scaling factors to convert them to 28 nm technology as shown in~\cite{stillmaker2017scaling}. 
The energy of HBM 1.0 is estimated with 7 pJ/bit as in~\cite{o2014highlights}.

\textbf{Benchmark LLM Datasets.} We conduct comprehensive experiments on the Mamba, which are owing to critical and efficient influence in recent model advancements. 
We depict Mamba models with different size and hyperparameters as shown in Table~\ref{tab:mamba}.
We focus on two primary metrics: perplexity and zero-shot performance.
The perplexity is evaluated by the WikiText~\cite{wikitext} and Lambada~\cite{lambada} benchmarks. 
The zero-shot performance is assessed across four zero-shot benchmarks, namely Piqa~\cite{piqa}, HellaSwag~\cite{hellaswag}, WinoGrande~\cite{winogrande}, and Arc-easy~\cite{arc-ec}. 

\textbf{Baseline Platform.} 
To compare the performance and energy consumption of \we with state-of-the-art works, we evaluate Mamba model on a Linux workstation equipped with one Intel Xeon 8358P CPU~\cite{intel_8358p} and a 252 GB DDR4 memory and one NVIDIA Tesla A100 GPU~\cite{a100}, denoted as Mamba-CPU and Mamba-GPU, respectively. 
Table~\ref{tab:system_config} lists the system configurations for above implementations.

\begin{table}[htbp]
\small
    \centering
    \vspace{-10pt}
    \caption{Hyperparameters of Models in Mamba Family}
    \vspace{-12pt}
    \begin{tabular}{cccccc}
        \toprule
        \textbf{Hyperparameters} & 130M & 370M & 790M & 1.4B & 2.8B \\
        \midrule
        \textbf{Layers} & 24 & 48 & 48 & 48 & 64 \\
        \textbf{Hidden Size} & 768 & 1024 & 1536 & 2048 & 2560 \\
        \bottomrule
    \end{tabular}
    \vspace{-12pt}
    \label{tab:mamba}
\end{table}

\begin{table}[t]
    \small
    \centering
    \caption{System Configuration.}
    \vspace{-12pt}
    \begin{threeparttable}
    \begin{tabular}{cccc}
        \toprule
         & \textbf{Mamba-CPU} & \textbf{Mamba-GPU} & \textbf{\we} \\
        \midrule
        \textbf{Compute} & 2.6GHz @  & 1.4GHz @ & 1GHz @ 32 RCUs\\
        \textbf{Unit} & 32 Cores & 8192+512 Cores & (each with 16$\times$16 RPEs)\\
        \midrule
        \textbf{On-chip} & \multirow{2}{*}{48MB} & \multirow{2}{*}{40MB} & \multirow{2}{*}{24MB}\\
        \textbf{Memory} & \multirow{2}{*}{} & \multirow{2}{*}{} & \\
        \midrule
        \textbf{Off-chip} & 136.5GB/s & 2039GB/s & 256GB/s\\
        \textbf{Memory} & DDR4 & HBM2e & HBM1.0\\
        \bottomrule
    \end{tabular}
    \vspace{-3pt}
    \begin{tablenotes}
    \small
        \item Note: GPU's on-chip memory includes the register files, and L1 and L2 caches. Mamba-GPU includes 8192 CUDA Cores and 512 Tensor Cores (each with 256 cores).
    \end{tablenotes}
    \vspace{-12pt}
    \end{threeparttable}
    \label{tab:system_config}
\end{table}


\subsection{Accuracy Evaluations}
Table \ref{tab:nonlinear loss} illustrates the perplexity and zero-shot performance of the approximation algorithm on Mamba families, compared to the metrics computed using the original approach, along with the accuracy of the original fast exp algorithm applied to exponential and SiLU computations. The experimental results indicate that benefiting from the bias introduced based on the data distribution, our improved algorithm outperforms the fast exp algorithm across all sizes of Mamba. The average accuracy improvement ranges from 0.19\% to 0.44\%. The difference in accuracy compared to the original algorithm does not exceed 0.29\%, indicating minimal loss in precision. 
After incorporating the piecewise SiLU into our complete algorithm, the maximum accuracy loss is only 0.84\% . While employing the fast exp algorithm in SiLU yields higher accuracy, it introduces divide unit into PE unit, resulting in significant area overhead.


\begin{table}[!t]
\small
    \caption{Perplexity and Accuracy under varied approximation algorithms}
    \centering
    \vspace{-12pt}
    \label{tab:nonlinear loss}
    \begin{tabular}{lccc}
    \toprule
        \multirow{2}{*}{Method} & Perplexity ($\downarrow$) & \multicolumn{2}{c}{Accuracy ($\uparrow$)}\\
        \cmidrule(lr){2-4}
        ~ & Wikitext/Lambada & Piqa/Wino./Arc-E/Hella. & Avg.($\uparrow$) \\ 
        \midrule
        Mamba-130M & 26.25/16.04 & 63.17/52.33/42.09/35.23 & 48.21 \\
        fast\_exp & 49.61/300.56 & 63.82/50.75/41.33/34.97 & 47.71 \\
        \textbf{Our\_exp} & 27.36/19.49 & 63.22/51.62/41.84/35.01 & 47.92 \\
        \textbf{Our\_silu} & 28.58/18.95 & 63.60/51.54/41.54/35.63 & \textbf{48.08} \\
        \textbf{Ours} & 29.69/18.50 & 63.93/51.54/40.87/35.46 & 47.95 \\
        \midrule
        Mamba-370M & 18.25/8.14 & 68.28/55.41/48.15/46.46 & 54.58 \\
        fast\_exp & 29.46/136.70 & 68.72/55.25/47.81/45.36 & 54.28 \\
        \textbf{Our\_exp} &  18.77/7.86 & 68.72/55.49/47.52/45.99 & 54.43 \\
        \textbf{Our\_silu} & 20.07/8.62 & 68.55/55.88/47.14/47.69 & \textbf{54.82} \\
        \textbf{Ours} & 20.65/8.29 & 69.15/55.41/46.93/47.25 & 54.69 \\
        \midrule
        Mamba-790M & 15.06/6.01 & 72.58/55.64/53.83/55.05 & 59.28 \\
        fast\_exp & 23.45/92.31 & 72.31/55.80/54.59/54.22 & 59.23 \\
        \textbf{Our\_exp} & 15.39/6.60 & 72.47/56.59/54.50/54.48 & 59.51 \\
        \textbf{Our\_silu} & 16.10/6.42 & 71.22/58.25/52.57/55.63 & 59.42 \\
        \textbf{Ours} &  16.64/6.53 & 72.47/57.70/52.69/55.61 & \textbf{59.62} \\
        \midrule
        Mamba-1.4B & 13.57/5.04 & 73.88/61.17/61.15/59.14 & 63.83 \\
        fast\_exp & 20.66/58.84 & 73.45/60.77/60.23/58.29 & 63.18 \\
        \textbf{Our\_exp} & 13.83/5.45 & 73.83/60.62/61.28/58.76 & \textbf{63.62} \\
        \textbf{Our\_silu} & 14.71/5.92 & 73.88/59.75/59.09/59.28 & 63.00 \\
        \textbf{Ours} & 15.21/6.06 & 73.99/60.22/58.67/59.07 & 62.99 \\
        \midrule
        Mamba-2.8B & 11.76/4.23 & 75.79/63.38/64.27/66.17 & 67.40 \\
        fast\_exp & 17.51/37.66 & 75.79/63.54/64.35/65.14 & 67.20 \\
        \textbf{Our\_exp} & 11.98/4.52 & 75.57/63.38/65.19/65.40 & \textbf{67.39} \\
        \textbf{Our\_silu} & 12.72/5.86 & 75.03/63.22/61.99/65.67 & 66.48 \\
        \textbf{Ours} & 13.12/6.08 & 75.63/64.25/60.56/65.91 & 66.59 \\
        \bottomrule
    \end{tabular}
\vspace{-20pt}
\end{table}

 \begin{figure*}[!t]
  \centering
  \vspace{-7pt}
  \includegraphics[width=0.99\textwidth]{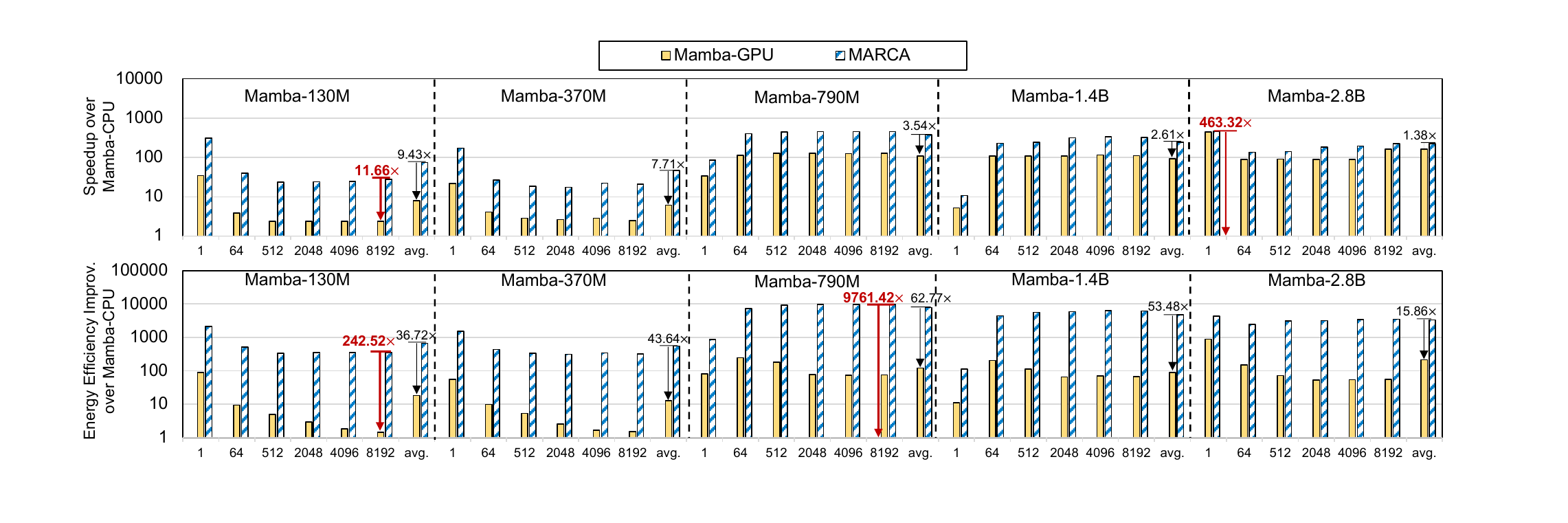}
  \vspace{-10pt}
  \caption{Comparison of speedup and energy efficiency improvement to Mamba-CPU and Mamba-GPU on Mamba models with different input sequence length.}
  \vspace{-13pt}
  \label{fig:speedup}
\end{figure*}

\subsection{Hardware Evaluations}
We compare our work with Mamba-CPU and Mamba-GPU in terms of speedup and energy consumption,
Finally, the area and power of our design is presented.

\textbf{Speedup.}
Figure~\ref{fig:speedup} top depicts that \we achieves up to 463.32$\times$/11.66$\times$ speedup and average 194.26$\times$/4.93$\times$ speedup compared with Mamba-CPU and Mamba-GPU, respectively.
The performance improvement comes from the reconfigurable architecture, and the intra-operation and inter-operation buffer management strategy. 
First, the reconfigurable computing unit with several reconfigurable processing element arrays accelerate the computations. 
Second, the intra-operation input data sharing and inter-operation data sharing methods reduce large redundant data between global memory and on-chip memory.

\textbf{Energy Efficiency.}
As Figure~\ref{fig:speedup} bottom shows, \we improves up to 9761.42$\times$/242.52$\times$ and average 3415.55$\times$/42.49$\times$ energy efficiency compared to Mamba-CPU and Mamba-GPU, respectively. 
We consider the energy consumption of all platforms includes the off-chip memory.

\textbf{Power and Area.}
The total power and area of \we are only 10.44 $W$ and 221.88 $mm^2$, respectively. 
For the on-chip buffer, we use eDRAM to reduce both the area and energy consumption. 
For the computation precision, we use 32-bit fixed point that is enough to maintain the accuracy of Mamba inference. 
Table~\ref{tab:pwr_and_area} provides area and power breakdown.
The on-chip buffer consumes most of power (>60\%) and area ($\sim$80\%) to support more memory access and data sharing for element-wise operations.
The compute engine consume 38.67\% power and 20.57\% area to perform the linear, element-wise, exponential, and SiLU computations.
The others are small owing to the simple implementations of normalization, and instruction fetching and decoding.

\begin{table}[htbp]
\small
    \centering
    \vspace{-10pt}
    \caption{Layout Characteristics of \we}
    \vspace{-12pt}
    \begin{tabular}{cccc}
        \toprule
        \textbf{Component} & \textbf{Sub-module} & \textbf{Area} ($mm^2$/\%) & \textbf{Power} ($W$/\%) \\
        \midrule
        Inst. Processing & - & 0.45/0.20\% & 0.045/0.43\%\\
        \midrule
        Norm. Unit & - & 0.06/0.03\% & 0.003/0.03\%\\
        \midrule
        \multirow{4}{*}{Compute Engine} & RPEs & 44.87/20.22\% & 3.92/37.55\% \\
        & Reduction Trees & 0.47/0.21\% & 0.053/0.51\% \\
        & Control Unit & 0.32/0.14\% & 0.064/0.61\% \\
        \cmidrule(lr){2-4}
        & - & 45.66/20.57\% & 4.037/38.67\% \\
        \midrule
        On-chip Buffer & - & 175.71/79.19\% & 6.35/60.87\%\\  
        \midrule
        Total & - & 221.88/100\% & 10.44/100\%\\ 
        \bottomrule
    \end{tabular}
    \vspace{-12pt}
    \label{tab:pwr_and_area}
\end{table}

\subsection{Ablation Study}

\begin{figure}[!t]
  \centering
  \includegraphics[width=0.47\textwidth]{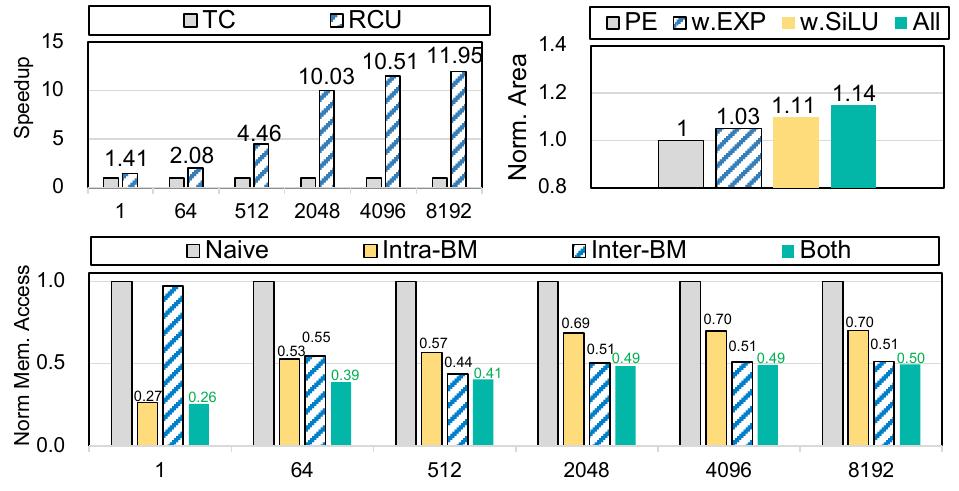}
  \vspace{-15pt}
  \caption{Ablation study: (1) Speedup of RCU over Tensor Core. (2) Normalized area of RPE. (3) Normalized memory access of intra-/inter-operation buffer management strategy.}
  \vspace{-15pt}
  \label{fig:ablation}
\end{figure}

\textbf{Speedup of RCU over Tensor Core.}
Figure~\ref{fig:ablation} top left shows the speedup from 1.41$\times$ to 11.95$\times$ over Tensor Core-based architecture with different sequence length in Mamba. 

\textbf{Normalized area of RPE.}
In Figure~\ref{fig:ablation} top right, we compare the normalized PE area overhead by supporting different nonlinear functions.
It shows that our reusable RPE only increases 14\% area overhead.

\textbf{Normalized memory access improvement.}
Figure~\ref{fig:ablation} bottom reveals the normalized global memory access with out buffer management strategy. When the sequence is short, linear operations are dominant, the intra-BM reduces 73\% memory access significantly. The inter-BM reduces 49\% memory access with long sequence.

\section{Conclusions}\label{sec:conclusion}
Our \we is the first proposed accelerator with reconfigurable architecture specifically tailored for Mamba computations.
We propose a reduction alternative PE array architecture to support both linear and element-wise operations. 
Then, based on the reconfigurable PE, we decompose the nonlinear functions and reuse PE arrays reduce the area overhead.
We also propose intra-operation and inter-operation buffer management strategy to maximize data reuse for two dominant operations.
We conduct extensive experiments on Mamba model families with different model sizes.
\we achieves up to 463.22$\times$/11.66$\times$ speedup and up to 9761.42$\times$/242.52$\times$ energy efficiency compared to Intel Xeon 8358P CPU and NVIDIA Tesla A100 GPU implementations, respectively.

\section{Acknowledgments}
This work was supported by the National Natural Science Foundation of China (No. 62104128, U21B2031), Beijing Douyin Information Service Co., Ltd.

\bibliographystyle{ACM-Reference-Format}
\bibliography{refs}

\end{document}